\documentclass[12pt]{iopart}

\usepackage{epsfig}
\usepackage{multirow}
\usepackage{iopams}

\begin{document}

\vspace*{-2cm}

\centerline{\large {\bf A simplified model of the source channel of}}
\centerline{\large {\bf the Leksell GammaKnife$^{\circledR}$ tested
with PENELOPE}}

\vspace{.3cm}

\begin{center}
{\bf Feras M.O. Al-Dweri and Antonio M. Lallena}\\
Departamento de F\'{\i}sica Moderna, 
Universidad de Granada, E-18071 Granada, Spain.

{\bf Manuel Vilches}\\
Servicio de Radiof\'{\i}sica, Hospital Cl\'{\i}nico
``San Cecilio'', Avda. Dr. Ol\'oriz, 16, E-18012 Granada, Spain.
\end{center}

\vspace{.2cm}

\small{
Monte Carlo simulations using the code PENELOPE have been performed to
test a simplified model of the source channel geometry of the Leksell
GammaKnife$^{\circledR}$. The characteristics of the radiation passing
through the treatment helmets are analysed in detail. We have found
that only primary particles emitted from the source with polar angles
smaller than 3$^{\rm o}$ with respect to the beam axis are relevant
for the dosimetry of the Gamma Knife. The photons trajectories
reaching the output helmet collimators at $(x,y,z=236~{\rm mm})$, show
strong correlations between $\rho=(x^2+y^2)^{1/2}$ and their polar
angle $\theta$, on one side, and between $\tan^{-1}(y/x)$ and their
azimuthal angle $\phi$, on the other. This enables us to propose a
simplified model which treats the full source channel as a
mathematical collimator. This simplified model produces doses in
excellent agreement with those found for the full geometry. In the
region of maximal dose, the relative differences between both
calculations are within 3\%, for the 18 and 14~mm helmets, and 10\%,
for the 8 and 4~mm ones. Besides, the simplified model permits a
strong reduction (larger than a factor 15) in the computational time.
}

\vspace{.3cm}

\section{Introduction}

Leksell GammaKnife$^{\circledR}$ (GK) is an instrument that permits a
precise external irradiation to treat intracraneal lesions.
Concentrating the radiation coming from 201 $^{60}$Co sources, a high
dose can be delivered to the target area with an accuracy better than
0.3~mm (Wu \etal 1990, Elekta 1992, Benjamin \etal 1997). Besides, the
sharp dose gradient allows to reduce the doses absorbed by the
critical brain structures surrounding the lesion to be treated. The
size and shape of the final beam can be fixed by combining different
treatment helmets with an adequate configuration of plugged and
unplugged sources. In this way the isodose distribution curves can be
modified in the optimal way. The GK is used together with
GammaPlan$^{\circledR}$ (GP), a planning system which uses
semi-empirical algorithms with various approximations to calculate
the doses, and assumes that the target is composed of unit density
material (Wu \etal 1990, Wu 1992, Elekta 1996).

To ensure the quality of the planning system, an experimental
verification is needed (Hartmann \etal 1995), but the difficulties
inherent to the physical dose measurements make Monte Carlo (MC)
simulations to be a good complementary tool to achieve this
purpose. In the case of the GK, a number of such calculations have
been performed in the last few years. Most of them have been done with
EGS4 (Cheung \etal 1998, 1999a, 1999b, 2000, Xiaowei and Chunxiang
1999) and have shown no relevant differences from GP
calculations for a homogeneous phantom. However, in stereotactic
radiosurgery, Solberg \etal (1998) have pointed out a remarkable
disagreement between MC results and those predicted by the usual
planning systems, once phantom inhomogeneities are taken into
account. For the GK, Cheung \etal (2001) have found discrepancies up
to 25\% in case of extreme conditions (mainly near tissue interfaces
and dose edges.)

Recently, Moskvin \etal (2002) have performed a detailed study in
which they have determined the characteristics of the beams emitted
after the helmets and have used them to calculate the dose field
inside a polystyrene phantom. These authors have used PENELOPE
(v.~2000) and have found a good agreement with the calculations
performed in previous works and with the predictions of the GP, what
ensures the suitability of PENELOPE for this kind of simulations.

In the present work we want to go deeper in some of the basic aspects
analysed by Moskvin and collaborators (2002). We have simulated a
single source of the GK using the version 2001 of PENELOPE (Salvat
\etal 2001) and we have investigated the energy spectra, the particle
spatial distributions and the correlations between the polar and
azimuthal angles of the particle trajectories and the coordinates of
the point these trajectories reach the treatment helmets. The results
we have obtained permit a simplification of the full geometry which
allows a considerable reduction of the simulation CPU time, without
loss of accuracy in the doses delivered to the phantom.

In the next section we describe the geometrical model we have used to
simulate the single source of the GK as well as the relevant details
concerning the MC calculations we have performed. Section 3 is devoted
to discuss the results we have obtained and study the geometrical
simplification we propose. Finally we draw our conclusions.

\section{Material and Methods}

\subsection{Leksell GammaKnife$^{\circledR}$ model}

\begin{figure}[ht]
\begin{center}
\epsfig{figure=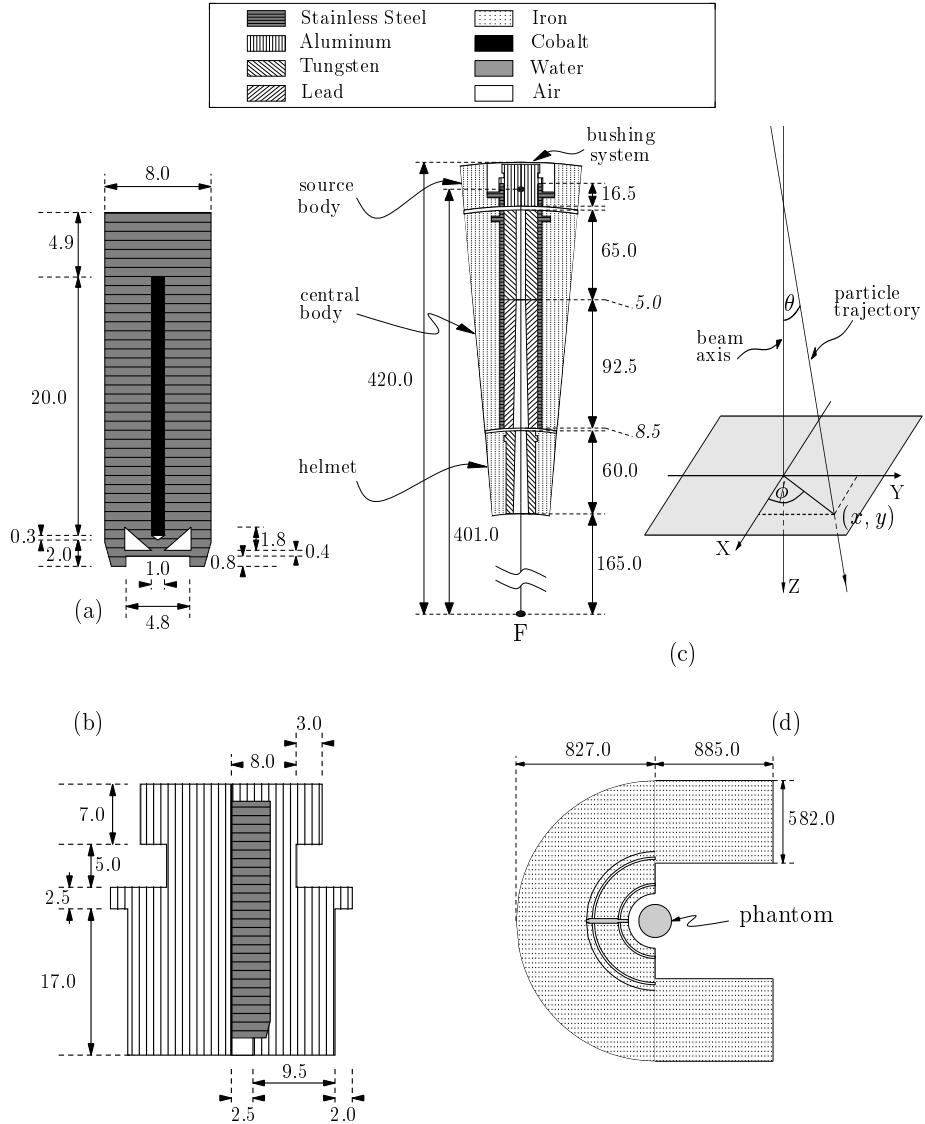,width=13cm}
\end{center}
\caption{Schematic view of the geometrical models used in our
simulations for the different parts of the GK: (a) capsule involving
the radionuclide; (b) bushing system surrounding the capsule; (c) full
collimation system for a single source, and (d) head of the GK showing
the situation of the water phantom. Distances are in mm. The values in
italic give the collimator apertures at the corresponding positions. F
labels the focus of the collimation system. In panel (c) the scheme
showing the polar and azimuthal angles and the $(x,y)$ coordinates
along the beam line is also included.
\label{fig:GK-scheme}}
\end{figure}

Details concerning the GK geometry have been obtained from the User
Manual (Elekta 1992). In actual simulations, the different pieces
forming the head and the collimating system have been described by the
specific geometries sketched in figure \ref{fig:GK-scheme}.

Figure \ref{fig:GK-scheme}a shows the capsule. It is made of stainless
steel (SS) and includes the $^{60}$Co active core. The latter (in black
in the figure) is formed by 20 cylindrical pellets of 1~mm diameter
and 1~mm height. In our simulations, the active core is considered as
an unique cylindrical source of 1~mm diameter and 20~mm height made of
cobalt.

The capsule is inside a bushing system made of aluminum. In our
simulations we have used the simplified geometry plotted in figure
\ref{fig:GK-scheme}b.

Figure \ref{fig:GK-scheme}c shows the geometry we have considered for
the collimation system of the single source. The radiation coming from
the $^{60}$Co source is collimated by a stationary collimator (located
in the central body) and an additional collimator located in the
helmet. The first one is made of tungsten and lead, while that of the
helmet is of tungsten. Four different helmets are available, producing
beams with nominal aperture diameters of 4, 8, 14 and 18~mm at the
focal point of the system, which is located at 401~mm from the source
centre. The inner and outer collimator diameters of the helmets we
have used are given in table \ref{tab:helmets} and have been taken
from Moskvin \etal (2002).

\begin{table}
\caption{
Inner and outer helmet collimator apertures of the
Leksell GammaKnife$^{\circledR}$. The values have been taken from
Moskvin \etal (2002).
\label{tab:helmets}}
\begin{center}
\begin{tabular}{@{}ccccc}
\br
Final beam diameter & 4~mm & 8~mm & 14~mm & 18~mm \\
\mr
Inner helmet collimator aperture [mm] & 2.0 & 3.8 & ~6.3 & ~8.3 \\
Outer helmet collimator aperture [mm]  & 2.5 & 5.0 & ~8.5 & 10.6 \\
\br
\end{tabular}
\end{center}
\end{table}

Finally, figure \ref{fig:GK-scheme}d shows the complete geometry
including the head of the instrument (which is made of iron) and the
phantom used to simulate the patient head. The latter is a sphere of
160~mm of diameter, filled with water and centered in the focus.
The situation of the single source channel we have considered
in our simulations is also shown.

\subsection{Monte Carlo calculations}

In this work we have used the PENELOPE (v. 2001) MC code (Salvat \etal
2001) to perform the simulations. PENELOPE is a general purpose MC
code which allows to simulate the coupled electron-photon
transport. It can be applied for energies ranging from a few hundred
eV up to 1~GeV, for electrons, photons and positrons and for arbitrary
materials. Besides, PENELOPE permits a good description of the
particle transport at the interfaces and presents an accurate
simulation at low energies. As an example we mention here that the
code has been used to simulate an accelerator head, obtaining results
in good agreement with measurements (Sempau \etal 2001).

In PENELOPE, analog simulation is performed for photons. On the other
hand, electron and positron simulation is done in a mixed scheme in
which collisions are classified in two types: hard and soft. Hard
collisions are simulated in a detailed way and are characterised by
polar angular deflections or energy losses larger than certain cutoff
values. Soft collisions do not fulfil these conditions and their
effects are described in terms of a condensed simulation based on a
multiple scattering theory. The electron tracking is controlled by
means of four parameters: $C_1$ and $C_2$, $W_{\rm cc}$ and $W_{\rm
cr}$. The first two refer to elastic collisions. $C_1$ gives the
average angular deflection produced by multiple elastic scattering
along a path length equal to the mean free path between consecutive
hard elastic events. $C_2$ represents the maximum value permitted for
the average fractional energy loss between consecutive hard elastic
events. On the other hand, $W_{\rm cc}$ and $W_{\rm cr}$ are energy
cutoffs to separate hard and soft events. Thus, the inelastic electron
collisions with energy loss $W<W_{\rm cc}$ and the emission of
Bremsstrahlung photons with energy $W<W_{\rm cr}$ are considered in
the simulation as soft interactions. The particles are simulated for
energies above a given absorption energy, $E_{\rm abs}$, below which
they are absorbed locally. Besides, the length of the steps generated
in the simulation is limited by an upper bound, $s_{\rm max}$. Table
\ref{tab:parameters} shows the values of these parameters for the
various materials used in our simulations. The values fixed for the
different parameters in the various materials permit to obtain results
in reasonable CPU times. The key point here concerns the relative
large absorption energies adopted for electrons in the heavier
materials, where high energy electrons produce a very large number of
secondary particles. These particles are not relevant for the
dosimetry in the phantom and the $E_{\rm abs}$ value we have chosen
avoid the loss of CPU time following their tracks. We have checked
that the use of the parameter set shown in table \ref{tab:parameters}
does not affect the precision of the final results.

\begin{table}
\caption{
Tracking parameters of various materials assumed in our PENELOPE
simulations. $E_{\rm abs}$($\gamma$) and $E_{\rm
abs}$(e$^{-}$,e$^{+}$) stand for the absorption energies corresponding
to photons and electrons and positrons, respectively.
\label{tab:parameters} }
\begin{center}
\begin{tabular}{@{}ccccc}
\br
   && \multicolumn{3}{c}{materials} \\
\mr
   & ~~ & Air & Water, SS, Al, Co & Fe, Pb, W \\ 
\mr
       $E_{\rm abs}$($\gamma$) [keV] && 1.0  & 1.0  & 1.0 \\
$E_{\rm abs}$(e$^{-}$,e$^{+}$) [keV] && 0.1  & 50.0 & 500.0 \\
                             $C_{1}$ && 0.05 & 0.1  & 0.1 \\
                             $C_{2}$ && 0.05 & 0.05 & 0.05 \\
                      $W_{\rm cc}$ [keV] && 5.0 & 5.0  & 5.0 \\
                      $W_{\rm cr}$ [keV] && 1.0 & 1.0  & 1.0 \\
                        $s_{\max}$ [cm]  && 
                            $10^{35}$ & $10^{35}$ & $10^{35}$ \\
\br
\end{tabular}
\end{center}
\end{table}                               

To finish with the description of the MC code, it is worth to mention
the main differences between the version 2001 that we have used and
the 2000, used by Moskvin \etal (2002), of PENELOPE. These affect the
model used to describe the electron and positron elastic collisions,
the Bremsstrahlung emission produced by electrons and positrons, the
photon photoelectric absorption cross sections and the fluorescence
processes taken into account. Details can be found in Salvat \etal
(2001). We have performed some basic simulations using both versions
of the code and the results obtained showed no differences. Then we do
not expect differences between our results and those of Moskvin \etal
(2002) other than those linked to possible differences in the
geometries considered.

The radionuclide was assumed to be uniformly distributed inside the
active core. $^{60}$Co decays via $\beta^-$ to excited
$^{60}$Ni. However, the emitted electrons do not exit the bushing
assembly. Excited $^{60}$Ni decays to its ground state by emitting two
photons with energies 1.17 and 1.33~MeV. In our simulations we have
assumed that photons are emitted with the average energy
1.25~MeV. This does not produce any effect on the results. The
direction of the emitted photons was supposed to be isotropic around
the initial position. In actual calculations, and in order to perform
the simulations in reasonable times, this direction was sampled in a
cone with a given semi-aperture $\theta_{\rm max}$. Then, $w=\cos
\theta_{\rm i}$, with $\theta_{\rm i}$ the initial polar angle of the
direction of the emitted photons, was sampled uniformly between 1 and
$\cos \theta_{\rm max}$ and the initial azimuthal angle $\phi_i$ was
sampled uniformly between 0 and $2\pi$.

The statistical uncertainties were calculated by scoring, in each
voxel, both the quantity of interest $Q$ and its square for each
history. Thus, the MC estimate of the quantity is given by
\[
 Q_{\rm mean} \, = \, \frac{1}{N} \sum_{i=1}^{N} q_{i} \, ,
\]
where $N$ is the number of simulated histories and $q_{i}$ is the
value scored by all particles of the $i$-th history (that is,
including the primary particle and all the secondaries it
generates). The statistical uncertainty is given by
\[
 \sigma_{Q_{\rm mean}} \, = \, 
 \sqrt{\frac{1}{N} \left[ \frac{1}{N}\sum_{i=1}^{N} q_{i}^2 
\, - \, Q_{\rm mean}^2 \right] } \, .
\]
If the contributions $q_i$ take the values 0 and 1 only, the
standard error can be evaluated as
\[
 \sigma_{Q_{\rm mean}} \, = \, 
 \sqrt{\frac{1}{N} \, Q_{\rm mean} \, (1 - {Q_{\rm mean}})} \, .
\]
The number of histories simulated has been chosen in each case to
maintain these statistical uncertainties under reasonable levels.
The uncertainties given throughout the paper correspond to
1$\sigma$.

\begin{table}
\caption{ 
Composition (in weight fraction) and densities of various
materials assumed in our PENELOPE simulations. 
\label{tab:materials}}
\begin{center}
{\scriptsize
\begin{tabular}{@{}ccccccccc}
\br
&\multicolumn{3}{c}{Compound Materials} &
 \multicolumn{5}{c}{Elementary Materials} \\ 
& {Air} & {SS} & {Water} 
& {Aluminum} & {Cobalt} & {Iron} 
& {Lead} & {Tungsten} \\
\mr
H  &          &         & 0.111894 & & & & & \\
C  & 0.000124 & 0.00026 &          & & & & & \\
N  & 0.755267 &         &          & & & & & \\
O  & 0.231781 &         & 0.888106 & & & & & \\
Al &          &         &          & 1.0 & & & & \\
Si &          & 0.0042  &          & & & & & \\
P  &          & 0.00019 &          & & & & & \\
S  &          & 0.00003 &          & & & & & \\
Ar & 0.012827 &         &          & & & & & \\
Cr &          & 0.1681  &          & & & & & \\
Mn &          & 0.014   &          & & & & & \\
Fe &          & 0.6821  &          & & & 1.0 & & \\
Co &          &         &          & & 1.0 & & & \\
Ni &          & 0.1101  &          & & & & & \\
Mo &          & 0.0211  &          & & & & & \\
W  &          &         &          & & & & & 1.0 \\
Pb &          &         &          & & & & 1.0 & \\
\mr
density [g cm$^{-3}$]
  & 0.0012048 & 7.8     & 1.0      & 
2.6989   & 8.9    & 7.874 & 11.35 & 19.3  \\
\br
\end{tabular}
}
\end{center}
\end{table}

The geometries discussed in the previous section were described by
means of the geometrical package PENGEOM of PENELOPE in terms of
quadric surfaces. Table \ref{tab:materials} gives the composition and
densities of the different materials used in our simulations.

The system for a single source presents cylindrical symmetry. The
reference system was located with the origin in the centre of the
active core, with the $z$ axis along the source and pointing toward
the phantom. The different magnitudes, and in particular the dose rate
$D(\rho,z)$, were supposed to be functions of the $z$ and $\rho
=(x^2+y^2)^{1/2}$ coordinates.

The full volume of the water phantom was subdivided into annular
volume bins with thicknesses $\Delta \rho=0.5$~mm and $\Delta z=1$~mm
in case of the 18 and 14~mm treatment helmets. For the smaller ones, 8
and 4~mm, the values considered were $\Delta \rho=0.25$~mm and $\Delta
z=0.5$~mm

\section{Results}

\subsection{Effective emission angle}

The first point to be fixed in order to perform the simulations is to
determine the value of the semi-aperture $\theta_{\rm max}$ of the
cone where the direction of the initial photon is sampled. To do that
we have defined the ``effective emission angle'', $\theta^\alpha_{\rm
eff}$, as the angle $\theta_{\rm i}$ of the direction of an emitted
primary photon such as it, or any of the secondary particles it
creates passing through the collimator, reaches a certain region
$\alpha$ (e.g. a body or a plane) in the simulation geometry.  In
particular we have studied the distribution of the effective emission
angle at the water phantom, $\theta^{\rm phantom}_{\rm eff}$.

\begin{figure}[ht]
\begin{center}
\epsfig{figure=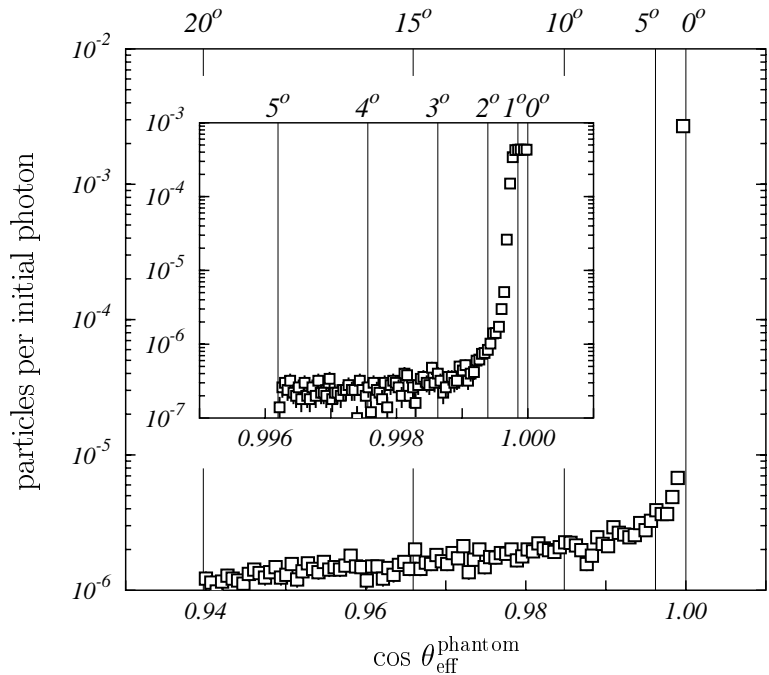,width=11cm}
\end{center}
\caption{Distribution of particles per initial photon vs. the cosine
of the effective emission angle at the water phantom, in the case of
the 18~mm helmet. A maximum emission angle $\theta_{\rm max}=20^{\rm
o}$ has been used in the simulations.
\label{fig:eff-ang}}
\end{figure}

Moskvin \etal (2002) sampled the initial directions of the emitted
photons in a cone with 10$^{\rm o}$ of semi-aperture and checked that
no differences, within statistical uncertainties, were observed by
increasing this value up to 90$^{\rm o}$ for the 18~mm helmet. We have
performed a simulation with the full geometry of the GK, for the 18~mm
helmet. A total of $6\cdot 10^7$ histories were followed and the
initial directions were sampled in a cone with semi-aperture
$\theta_{\rm max}=20^{\rm o}$. The distribution of the fraction of
initial photons vs. the cosine of $\theta^{\rm phantom}_{\rm eff}$ is
shown in figure \ref{fig:eff-ang}. As we can see, the distribution
reduces by three orders of magnitude between 0$^{\rm o}$ and 5$^{\rm
o}$. The insert in the same figure shows, with more detail, the
distribution in this interval. The number of counts for $\theta^{\rm
phantom}_{\rm eff}=2^{\rm o}$ ($3^{\rm o}$) is more than two (three)
orders of magnitude smaller than for $\theta^{\rm phantom}_{\rm eff}
\sim 0^{\rm o}$.

\begin{figure}[ht]
\begin{center}
\epsfig{figure=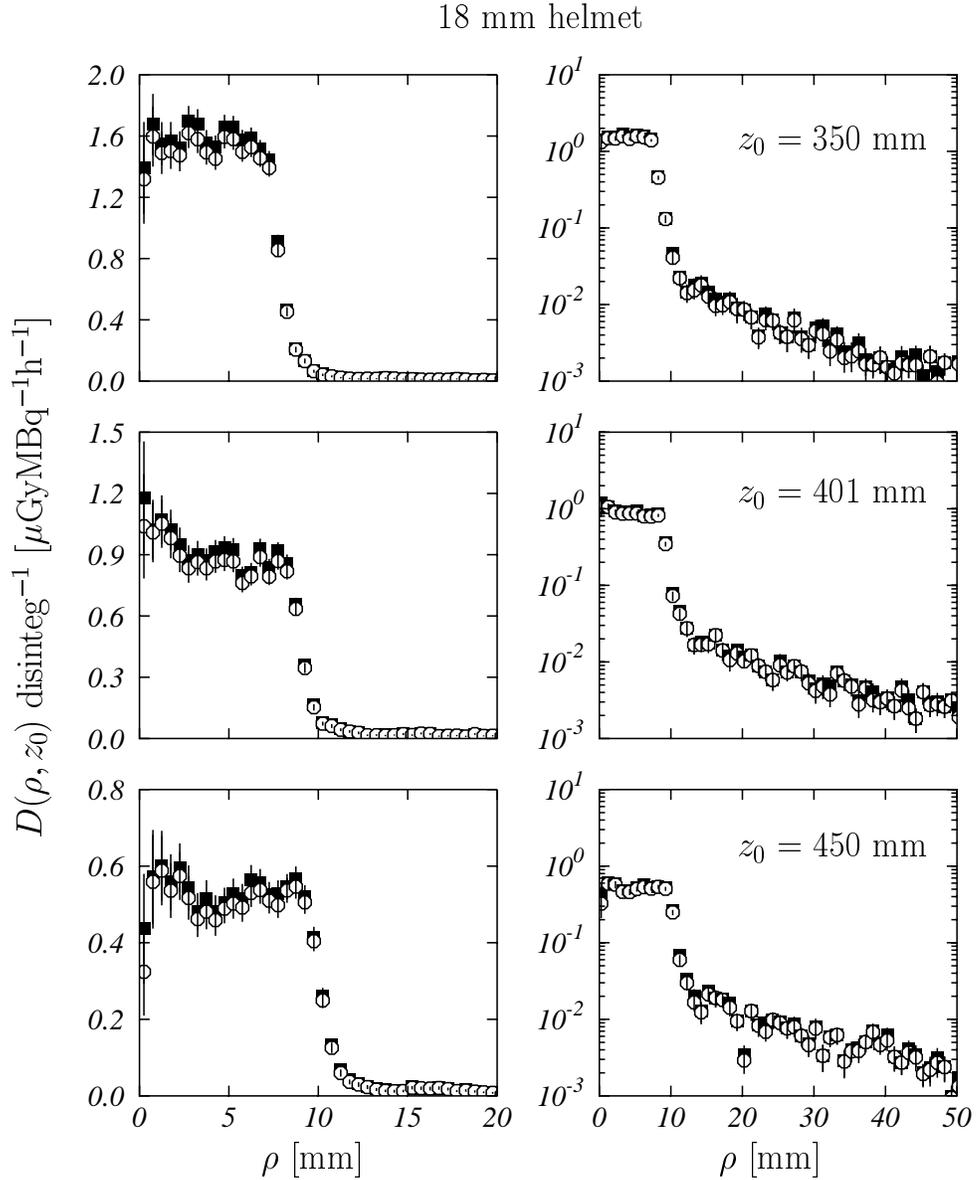,width=13cm}
\end{center}
\caption{Dose profiles (per disintegration) for the 18~mm helmet as a
function of the radial distance from the beam axis $\rho$. The results
obtained at the planes $z_0=350$, 401 and 450~mm are shown. The
simulations have been performed for $\theta_{\rm max}=20^{\rm
o}$. Black squares correspond to the dose deposited in all the
histories followed, while open circles show the dose deposited in
those histories in which the primary photons were emitted with
$\theta_{\rm i} \leq 3^{\rm o}$ only. Left (right) panels are in
linear (semi-logarithmic) scale. The values include the geometrical
factor $(1-\cos 20^{\rm o})/2$ which renormalizes the number of
particles emitted in the $20^{\rm o}$ cone to the total number of
particles emitted in actual situations (which correspond to polar
angles between 0 and $\pi$).
\label{fig:dose-eff-ang}}
\end{figure}

Despite that a semi-aperture of 3$^{\rm o}$ for the cone in which the
initial directions are sampled is considerably smaller than that used
by Moskvin \etal (2002), we have checked that it suffices to obtain
right doses in the phantom.  We have performed a simulation in the
previous conditions, following $15\cdot 10^7$ histories. The dose
$D(\rho,z_0)$ per disintegration in planes transverse to the beam axis
direction with $z_0=350$, 401 and 450~mm (these are before the focus,
at the focus and after it, respectively) are plotted in figure
\ref{fig:dose-eff-ang}, as a function of the distance to the beam
axis, $\rho$. Therein black squares stand for the full calculation
(that is sampling the initial directions into the cone with
semi-aperture 20$^{\rm o}$), while open circles give the dose due to
histories whose the primary particle was emitted with an angle
$\theta_{\rm i} \leq 3^{\rm o}$. Both linear (left) and logarithmic
(right) plots are included to see the small and large $\rho$ behaviour
clearly. As we can see both results coincide within the statistical
uncertainties and, as a consequence, we can state that a maximum polar
angle $\theta_{\rm max}=3^{\rm o}$ for sampling the initial particle
direction is enough to ensure the accuracy of the results.

\subsection{Comparison with previous results}

With the value $\theta_{\rm max}=3^{\rm o}$ determined as explained in
the previous section, we have performed simulations considering the
full geometry of the GK and following $6\cdot 10^7$ histories.

\begin{figure}[ht]
\begin{center}
\epsfig{figure=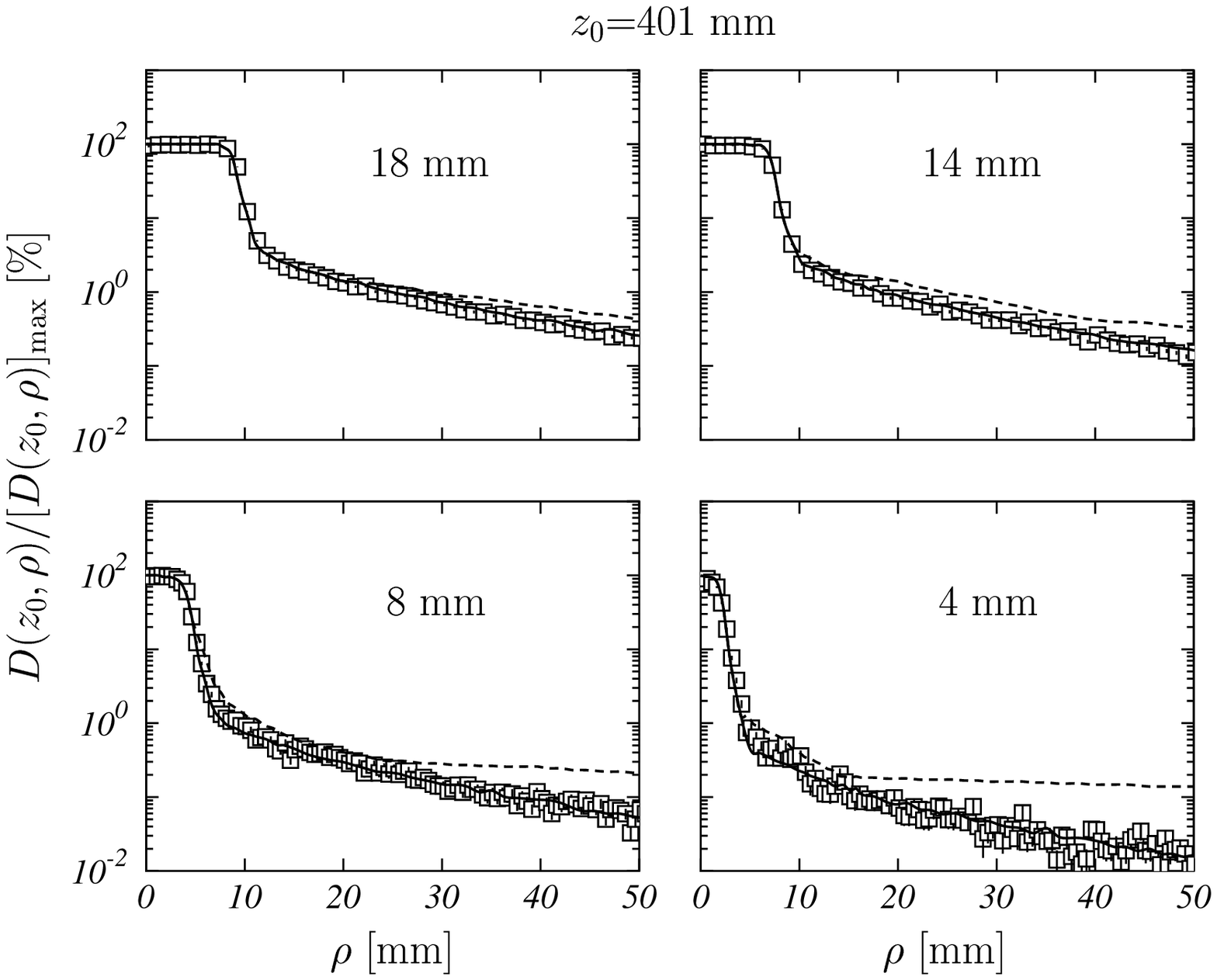,width=13cm}
\end{center}
\caption{Dose profiles (relative to their maximum) as a function of
the radial distance from the beam axis $\rho$. The results obtained in
our simulations at the plane $z_0=401$~mm for the four helmets 
(open squares) are compared with the results of Cheung \etal (1998)
(solid lines) and with the data provided by the manufacturer
(broken curves).
\label{fig:dose-tr1}}
\end{figure}

Figure \ref{fig:dose-tr1} shows the dose profiles in the plane of the
focus, transverse to the beam axis ($z=401$~mm) for the four helmets.
Therein the squares correspond to the simulations we have performed,
normalised to the maximum. Full curves show the results of Cheung {\it
et al.} (1998), while broken curves are given by the
manufacturer.\footnote{These two sets of results have been obtained by
scanning directly figures 1-4 of Cheung \etal (1998).} The agreement
with the results of Cheung \etal (1998) is very good (these authors
quoted an error below 1\% at dose maximum), while a clear discrepancy
with data provided by the manufacturer appears in the tail of the
distributions.

\begin{table}[hb]
\caption{ 
Values of $\rho_{50}$ and penumbra ($\rho_{80}-\rho_{20}$) obtained
for the four helmets and compared with the results quoted in Wu \etal
(1990). $\rho_{80}$, $\rho_{50}$ and $\rho_{20}$ stand for the $\rho$
distances providing, respectively, the 80, 50 and 20\% of the maximum
dose in the plane of the focus, transversse to the beam axis
($z=401$~mm).
\label{tab:penumbras} }
\begin{center}
\begin{tabular}{@{}cccccc}
\br
& \multicolumn{2}{c}{This work} &~& \multicolumn{2}{c}{Wu \etal
 (1990)} \\ \cline{2-3} \cline{5-6}
& ~~$\rho_{50}$ [mm]~~ & $\rho_{20}-\rho_{80}$ [mm] &&
  ~~$\rho_{50}$ [mm]~~ & $\rho_{20}-\rho_{80}$ [mm] \\ 
\mr
~4~mm & 2.0$\pm$0.3 & 1.1$\pm$0.4 && 2.0 &{1-2}\\ 
~8~mm & 4.2$\pm$0.3 & 1.4$\pm$0.4 && 4.2 &\\ 
14~mm & 7.2$\pm$0.5 & 1.4$\pm$0.7 && 7.0 &\\ 
18~mm & 9.1$\pm$0.5 & 1.3$\pm$0.7 && 9.0 &  \\ 
\br
\end{tabular}
\end{center}
\end{table}

Table \ref{tab:penumbras} quotes the values we have obtained for the
$\rho$ distance providing 50\% of the maximum dose, $\rho_{50}$, and
the penumbra, determined as the difference $\rho_{80}-\rho_{20}$
between the distances providing 80 and 20\% of the maximum dose. In
this table our findings are compared with those of Wu \etal (1990)
and, as we can see, the agreement is excellent.

\subsection{Characteristics of the beam after the helmets}

Now we analyse the characteristics of the beam after it goes through
the outer collimators of the helmets. We have performed simulations
with the full geometry of the GK, sampling the initial directions with
$\theta_{\rm max}=3^{\rm o}$ and following $6\cdot 10^7$ histories.

\begin{figure}[ht]
\begin{center}
\epsfig{figure=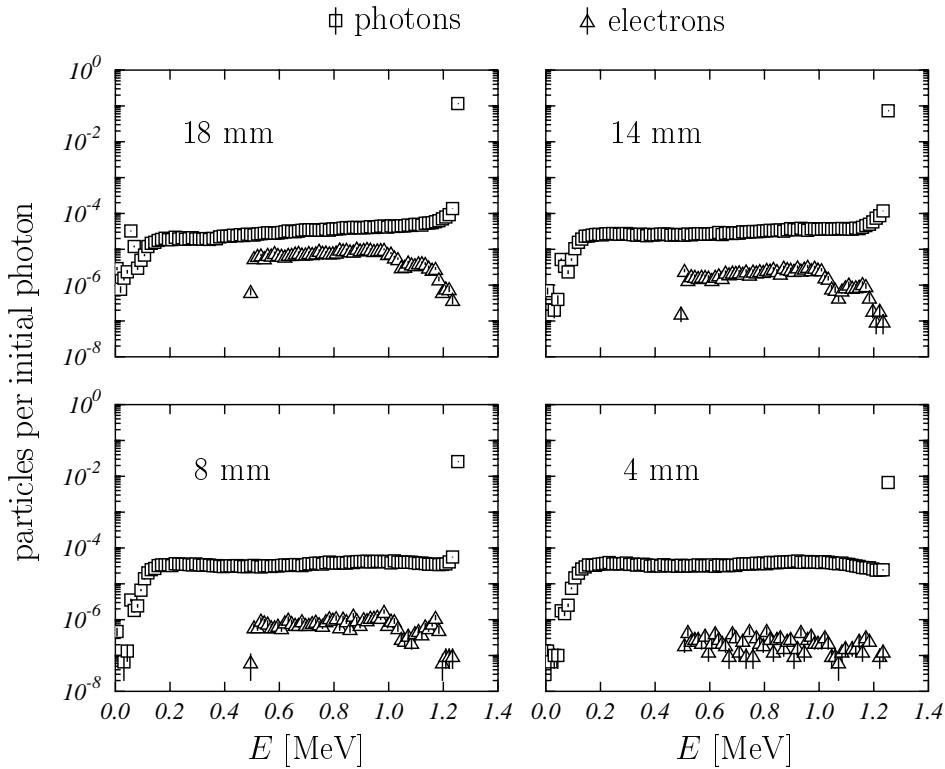,width=13cm}
\end{center}
\caption{Distributions of particles per initial photon vs. the energy
at the output collimators of the four treatment helmets. Squares
(triangles) correspond to photons (electrons).
\label{fig:energy-spectra}}
\end{figure}

First we have studied the energy spectra at the output collimators of
the four helmets and for both photons (squares) and electrons
(triangles). Results are plotted in figure \ref{fig:energy-spectra}.
In the four cases, the photon spectra show a pronounced
peak corresponding to the initial energy. For lower energies, the
spectra are rather uniform. Below 100 keV, the peaks due to
characteristic x rays emitted after the relaxation of vacancies
produced in K shells of the tungsten or lead atoms of the collimation
system can be distinguished, especially for the 18~mm helmet. In
general the number of particles exiting the output collimators of the
helmets diminishes, as expected, with their aperture.

In  what  respect  to  electrons,  one can  see  that  their  relative
importance reduces with the aperture of the helmet. On the other hand,
the artificial  discontinuity at 500~keV is due to the corresponding
absorption energy in the materials forming the collimators.

\begin{figure}[ht]
\begin{center}
\epsfig{figure=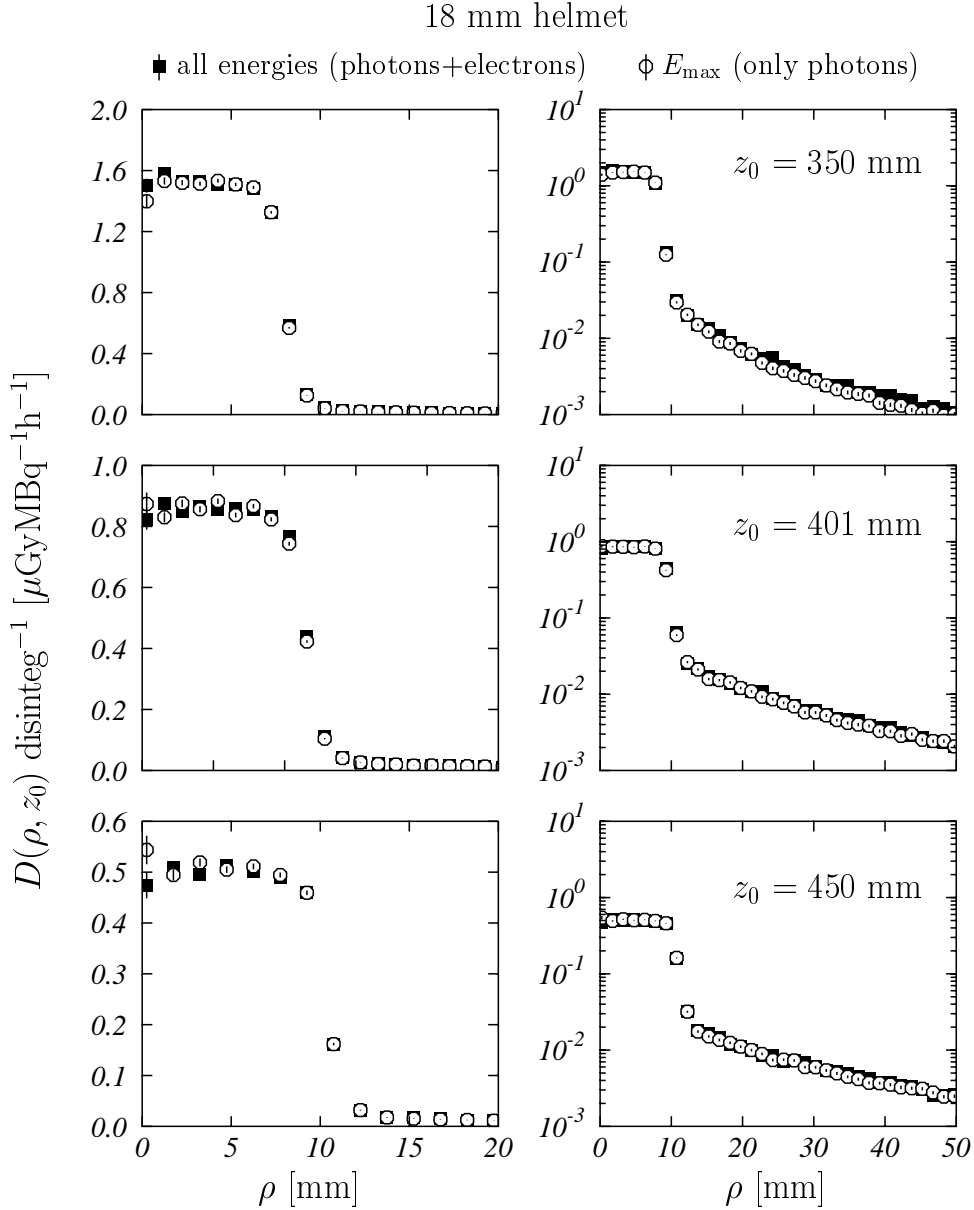,width=13cm}
\end{center}
\caption{Dose profiles (per disintegration) for the 18~mm helmet as a
function of the radial distance from the beam axis $\rho$. The results
obtained at the planes $z_0=350$, 401 and 450~mm are shown. The
simulations have been performed for $\theta_{\rm max}=3^{\rm
o}$. Black squares correspond to the dose deposited by photons and
electrons with any energy, while open circles give the dose due to
photons with the maximum energy only. Left (right) panels are in
linear (semi-logarithmic) scale. The values include the geometrical
factor $(1-\cos 3^{\rm o})/2$ which renormalizes the number of
particles emitted in the $3^{\rm o}$ cone to the total number of
particles emitted in actual situations (which correspond to polar
angles between 0 and $\pi$).
\label{fig:max-ener}}
\end{figure}

In view of these results, we can say that most of the particles
traversing the helmet output collimators have the initial energy
and, as a consequence, they are primary photons coming directly from
the source or, at most, having suffered a Rayleigh (elastic)
scattering. In figure \ref{fig:max-ener} we test the extent to which
the previous statement is correct for the 18~mm helmet. Therein we
have plotted the dose $D(\rho,z_0)$ per disintegration in the planes
$z_0=350$, 401 and 450~mm, as a function of the distance $\rho$ to the
beam axis. Left panels correspond to linear plots, while the right
ones are logarithmic plots pointing out the situation for large $\rho$
values. Black squares correspond to the full simulation, whereas open
circles are the results obtained by taking into account the
photons arriving to the phantom with the initial energy, only. As we
can see, the differences between both results are rather small.

\begin{figure}[ht]
\begin{center}
\epsfig{figure=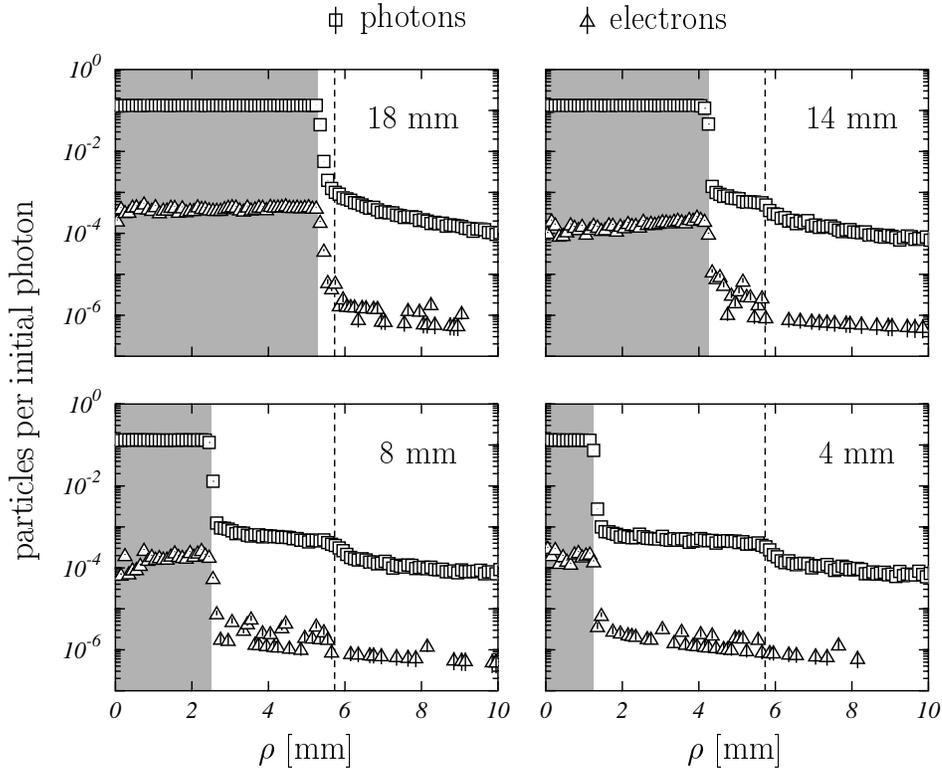,width=13cm}
\end{center}
\caption{Distributions of particles per initial photon vs. the 
radial distance from the beam axis $\rho$ at the output collimators of the
four treatment helmets. Shadow regions represent the semi-aperture of
these collimators. Squares (triangles) correspond to photons
(electrons). The dashed line indicates the semi-aperture of the beam
at the same $z$, if the helmets were not present.
\label{fig:dist-pos}}
\end{figure}

The second point of interest concerns the particle spatial
distribution at the helmet output collimators. Figure
\ref{fig:dist-pos} shows the particle distributions in the transverse
direction $\rho$ for the four treatment helmets. Squares (triangles)
correspond to photons (electrons). The results of the simulations are
divided by the number of initial histories and by the surface of the
scoring bin ($\pi (\rho_>^2-\rho_<^2)$), which has $\Delta
\rho=\rho_>-\rho_<=0.1$~mm. The shadow region indicates the
semi-aperture of the final collimators of the helmets.

The general trend is similar for the four helmets. The distribution
appears to be uniform inside the characteristic aperture, reducing
considerably for larger values of $\rho$. In case of the 14, 8 and
4~mm helmets, a second step at $\rho \sim 6$~mm is observed. This is
due to the output collimator of the central body of the collimation
system and corresponds, in fact, to the semi-aperture of the beam at
the same $z$, if the helmets were not present. The dashed line in
figure \ref{fig:dist-pos} indicates this semi-aperture. This step is
not observed in the case of the 18~mm helmet because it is hidden in
the sharp reduction of the distribution. Electrons show a similar
behaviour.

\begin{figure}[ht]
\begin{center}
\epsfig{figure=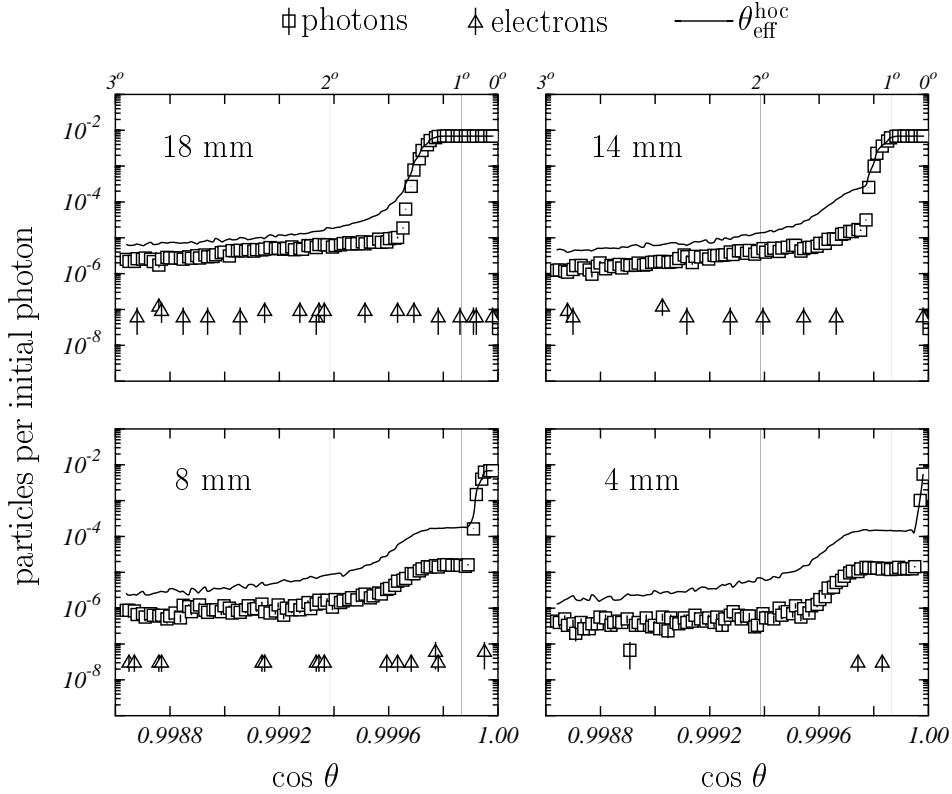,width=13cm}
\end{center}
\caption{Distributions of particles per initial photon vs. 
cos~$\theta$ for the trajectories traversing
the output collimators of the four treatment helmets. Squares
(triangles) correspond to photons (electrons). Solid lines represent
distributions of particles per initial photons vs.
the effective angle distributions at the same surfaces, 
$\theta^{\rm hoc}_{\rm eff}$.
\label{fig:dist-ang}}
\end{figure}

The particle distribution vs. the angle $\theta$ of the particle
trajectories once they pass through the helmet output collimators
deserves also to be analysed. Figure \ref{fig:dist-ang} shows with
squares (triangles) the distributions, per initial photon,
corresponding to photons (electrons). As we can see, in the case of
photons, there is a first region, for angles $\theta \sim 0^{\rm o}$,
where the distribution is uniform. In the case of the two smallest
helmets, a second region of uniformity appears for larger angles. The
first one is related to the characteristic aperture of each helmet
collimator, while the second is linked to the aperture of the final
collimator of the central body.

In the figure we have also plotted (full curves) the distributions of
the fraction of initial photons vs. the cosine of the effective
emission angle determined at the helmet outer collimators,
$\theta^{\rm hoc}_{\rm eff}$. These distributions are very similar to
the previous ones, the main difference being that those corresponding
to the output angles show a larger reduction after the first region of
uniformity. This points out the strong angular focalization produced
by the collimation system.

In what refers to the electron distributions, they show very low
statistics what indicates that their contribution to the dosimetry of
the GK is not relevant.

\subsection{Correlations}

An important aspect concerning the characteristics of the beams is
that associated to the correlations between the relevant variables,
these are, the energy of the particles, their positions and the angles
defining their trajectories once they reach the output helmet
collimators.

The study of the correlations of the energy with $\rho$ and $\theta$
shows that, as expected from the previous results, particles with
energies other than the initial one are practically not present,
irrespective of the $\rho$ or $\theta$ values one considers.

\begin{figure}[ht]
\begin{center}
\epsfig{figure=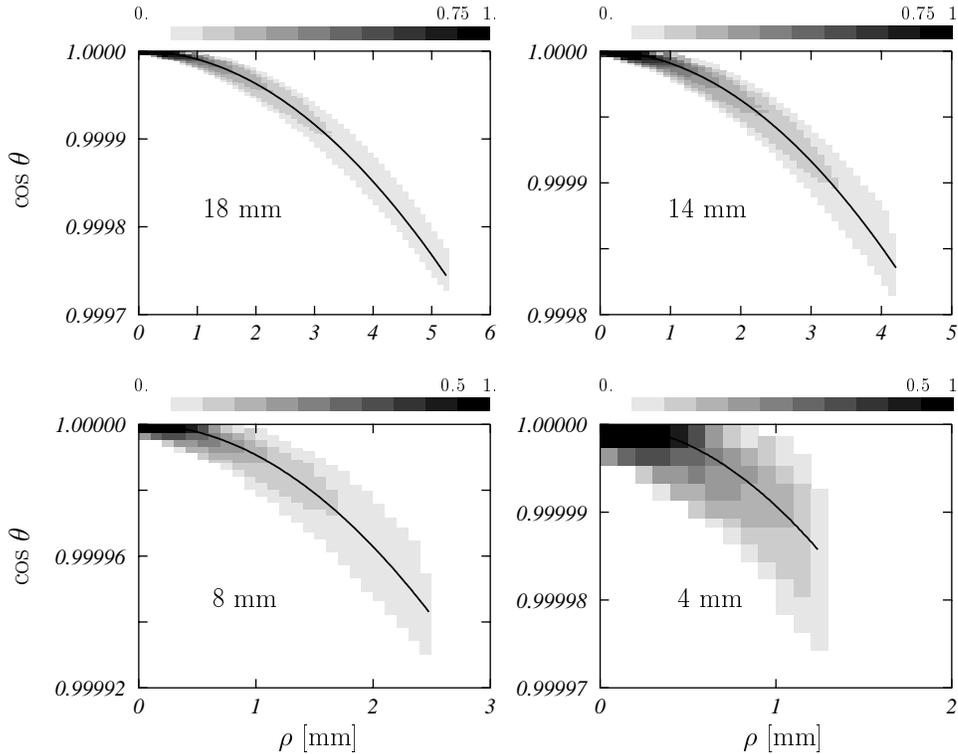,width=13cm}
\end{center}
\caption{Correlations between the $\rho$ distance to the beam centre
and $\cos \theta$ for the particle trajectories at the output
collimators of the four treatment helmets. Solid lines represent the
curves corresponding to the mathematical collimation considered in the
simplification described in the text.
\label{fig:the-rho1}}
\end{figure}

A more interesting result is shown by the correlations between the
geometrical variables. Figure \ref{fig:the-rho1} shows the results
found for $\rho$ and $\cos \theta$ and, as we can see, there exists a
strong correlation between both variables: the angle where the $\cos
\theta$ distribution is peaked increases with $\rho$. In this respect
it is worth to note that we have used a non-linear gray scale in order
to enhance the regions with lower correlations. These results agree
with the findings of Moskvin \etal (2002).

\begin{figure}[ht]
\begin{center}
\epsfig{figure=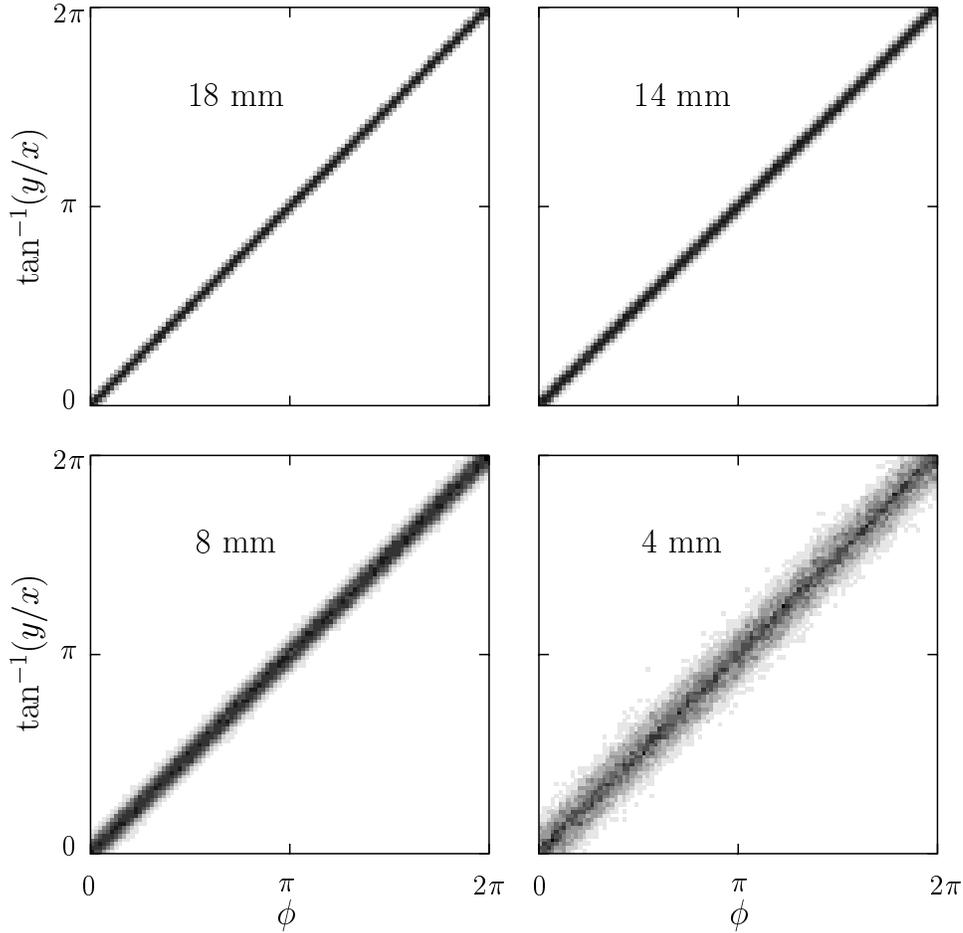,width=13cm}
\end{center}
\caption{Correlations between the azimuthal angle $\phi$ and
$\tan ^{-1} (y/x)$ for the particle trajectories at the output
collimators of the four treatment helmets. 
\label{fig:phy-xy}}
\end{figure}

However, these are not the unique correlations between the geometrical
variables involved in the problem. Figure \ref{fig:phy-xy} represents
the correlations shown by the azimuthal angle of the trajectory of the
particle and the quantity $\tan^{-1}(y/x)$ calculated with the
coordinates of this trajectory at the output helmet collimators. As we
can see, these two quantities are strongly correlated for the four
treatment helmets. The larger spread shown by the helmets with the
smaller collimators are due to the lower statistics in these cases.
This correlation was not explicitly mentioned by Moskvin \etal
(2002), but, however, it is on the basis of the 
simplification of the source channel geometry we discuss below.

\begin{figure}[ht]
\begin{center}
\epsfig{figure=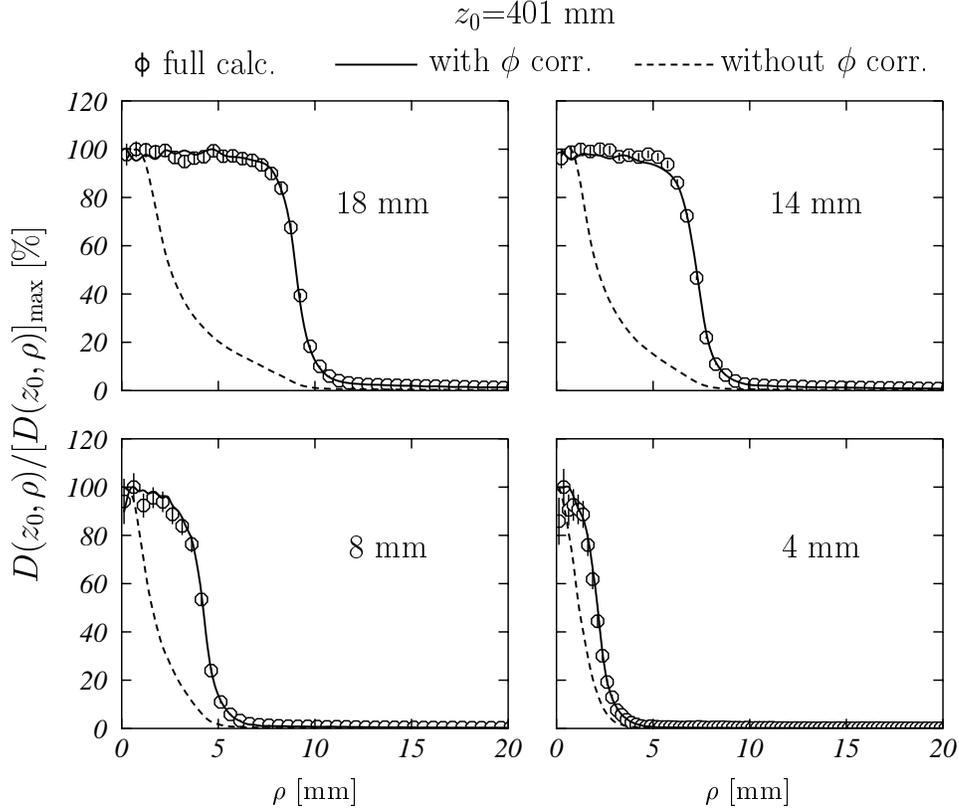,width=13cm}
\end{center}
\caption{Dose profiles (relative to their maximum) as a function of
the radial distance from the beam axis $\rho$. The results obtained in
our simulations at the plane $z_0=401$~mm for the four helmets are
shown. Open circles correspond to our full
calculation. Solid lines have been obtained by taking into account the
correlations found for the different variables (see text). Dashed
lines correspond to a similar calculation but ignoring the $\phi$
angle correlations, that is by sampling the $\phi$ angle uniformly
between 0 and 2$\pi$.
\label{fig:dose-phicor}}
\end{figure}

To test these correlations we have found, we have performed a new
simulation in which the particles are emitted from the helmet outer
collimators, according to the distributions discussed above and
conditioned to these correlations.  The procedure we have carried out
has been the following. First, we have sampled the initial position
$(x,y)$ uniformly within the aperture of the helmet outer
collimators. Second, the value of $\cos \theta$ for the trajectory has
been sampled according to the conditional distribution for the
corresponding $\rho$ value (see figure \ref{fig:the-rho1}). Finally,
the azimuthal angle $\phi$ has been sampled according to the
conditional distribution corresponding to the particular value
$\tan^{-1}(y/x)$ (see figure \ref{fig:phy-xy}). Only photons with
initial energy fixed to the maximum value 1.25~MeV have been
considered and a total of $2\cdot 10^7$ histories has been
followed. This strategy coincides with that used by Moskvin \etal
(2002). The results obtained in this way (solid lines) are compared to
those obtained with the full GK geometry (open circles) in figure
\ref{fig:dose-phicor}, normalized to the maximum. As we can see, the
agreement between both calculations is rather satisfactory.

In order to understand the implications of the correlations we have
encountered in the dosimetry of the GK, we have performed a new
simulation following the same procedure, but neglecting the
correlations shown by the azimuthal angle $\phi$, which has been
sampled uniformly between 0 an $2\pi$. The results obtained are
shown, again normalized to the maximum, with dashed lines in figure
\ref{fig:dose-phicor}. As we can see, if the $\phi$ angle correlations
are neglected, the shape of the dose profile at the focus is strongly
modified, showing a big enhancement at distances close to the beam
axis.

\subsection{Simplification of the source}

The simplification in the description of the source of the GK that we
propose in this work consists in substituting the full source channel
by a ``mathematical collimator'' in which a point source situated in
the center of the active core, emits photons with initial energy
equals to 1.25~MeV in the cone defined by itself and the outer
collimators of the treatment helmets.

The reasons for that rely, first, on the characteristics distributions
(energy, position and polar angle) of the particles traversing the
collimation channel, and, second and mainly, on the strong
correlations shown by $\phi$ and $\tan^{-1}(y/x)$ and by $\theta$ and
$\rho$, respectively. In fact, if we assume the
simplification mentioned, it is obvious that the azimuthal angle $\phi$
of a trajectory is related to the point $(x,y)$ at which it reaches the 
outer collimator source as (see the scheme in figure 1c) 
\[
\phi \, \equiv \, \tan^{-1}(y/x) \, ,
\]
a value which is in agreement with the maximum of the distributions in
figure \ref{fig:phy-xy}.  On the other hand, in this simplified
situation, the polar angle $\theta$ of this trajetory is related to
the corresponding value of $\rho$ in this surface as
\[
\tan \theta \equiv \displaystyle \frac{\rho}{z_{\rm col}} \, ,
\]
where $z_{\rm col}$ is the $z$ coordinate of the output collimator of
the helmets (236~mm in our case). In figure \ref{fig:the-rho1} we
have plotted, with solid lines, the quantity $\cos [\tan^{-1}
(\rho/z_{\rm col})]$ and, as we can see, it fits perfectly the maximum
of the respective distributions.

\begin{figure}[ht]
\begin{center}
\epsfig{figure=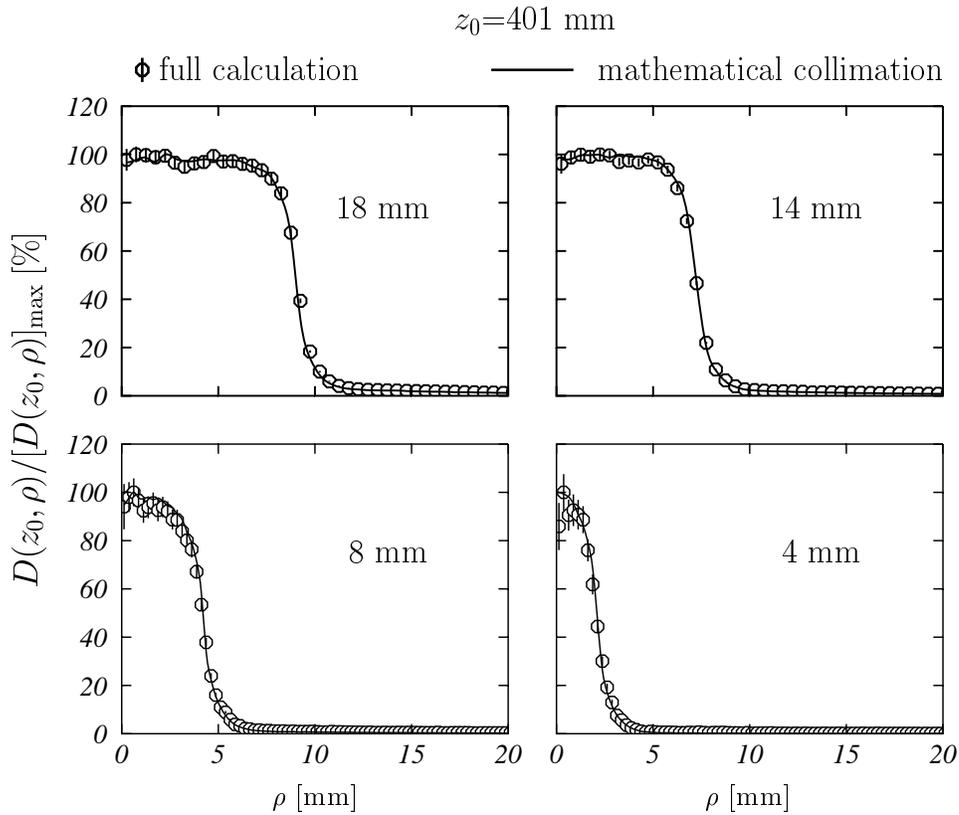,width=13cm}
\end{center}
\caption{Dose profiles (relative to their maximum) as a function of
the radial distance from the beam axis $\rho$. The results obtained in
our simulations at the plane $z_0=401$~mm for the four helmets are
shown. Open circles correspond to our
full calculation. Solid lines represent the calculation done with the
simplified model we have proposed.
\label{fig:dose-tr2}}
\end{figure}

In order to check the goodness of this simplification, we have
performed a new simulation following $2\cdot 10^7$ histories, assuming
a point source situated at the center of the $^{60}$Co source,
considering only photons with initial energy 1.25~MeV and performing a
``mathematical collimation'' by sampling the initial particle
directions isotropically inside the corresponding cones with apertures
equal to the diameters of the helmet outer collimators given in table
\ref{tab:helmets}. The results, normalized to the maximum, are plotted
with solid lines in figure \ref{fig:dose-tr2}, where we compare them
with those of the simulation considering the full GK geometry (open
circles). The uncertainties in the case of the simplified model are a
factor 3 (at least) smaller than those quoted for the full GK
geometry. The agreement between both simulations is remarkable. In
the region of maximal dose, the relative differences between both
calculations are within 3\%, for the 18 and 14~mm helmets, and 10\%,
for the 8 and 4~mm ones, where the uncertainties of the full
calculation results are relatively large. This shows the feasibility
of this simplification.

Besides, this simplified model provides a large reduction (more than a
factor 15) in the CPU time needed to perform the simulations.

\section{Conclusions}

In this work we have investigated the dosimetry of a single-source
configuration of the Leksell GammaKnife$^{\circledR}$ using the Monte
Carlo code PENELOPE (v. 2001). To do that a series of simulations have
been performed from which the following conclusions can be drawn: 
\begin{enumerate}
\item We have found that a maximum angle $\theta_{\rm max}=3^{\rm o}$
for sampling the initial direction of the emitted photons is enough to
ensure the accuracy of the results.

\item The characteristics of the beam after it goes trough the output
collimators of the helmets indicate that most of the particles
traversing these collimators are photons coming directly from the
$^{60}$Co source.

\item The polar angle $\theta$ and the distance to the beam axis
$\rho$ of the particle trajectories at the output collimators of the
helmets are strongly correlated. The same happens for the azimuthal
angle $\phi$ and the ratio $y/x$ of the corresponding coordinates.

\item These strong correlations can be explained by assuming that the
collimation system acts as a mathematical collimator in which the
particles are emitted in a cone defined by the output collimators of
the helmets and come from a point source located at the center of the
active core.
\end{enumerate}

The results obtained with the simplified geometry model we propose
here are in good agreement with those found for the full
geometry. Besides, the CPU time needed to perform the simulations is
largely reduced. This opens the possibility to use MC tools for
planning purposes in the GK. Obviously, a comparison of the results
obtained from calculations based on this simplified approach, for
various configurations of the 201 sources of the GK, with previous
works is necessary. Also it is worth to perform a detailed analysis of
the corresponding output factors. Work in these directions is being
done and will be published in a forthcoming paper.

\ack{ Authors wish to acknowledge G. Rey for useful discussion and for
providing us with geometrical details of the Leksell
GammaKnife$^{\circledR}$.  F.M.O. A.-D. acknowledges the
A.E.C.I. (Spain) and the University of Granada for funding his
research stay in Granada (Spain). This work has been supported in part
by the Junta de Andaluc\'{\i}a (FQM0225).}

\References

\item[] Benjamin W C, Curran W J Jr., Shrieve D C and Loeffler J S
1997 Stereotactic radiosurgery and radiotherapy: new developments and
new directions. {\it Semin. Oncol.} {\bf 24} 707-14

\item[] Cheung J Y C, Yu K N, Ho R T K and Yu C P 1999a Monte Carlo
calculations and GafChromic film measurements for plugged collimator
helmets of Leksell GammaKnife unit {\it Med. Phys.} {\bf 26} 1252-6
                             
\item[] Cheung J Y C, Yu K N, Ho R T K and Yu C P 1999b
Monte Carlo calculated output factors of a Leksell GammaKnife
unit
{\it Phys. Med. Biol.} {\bf 44} N247-9 

\item[] Cheung J Y C, Yu K N, Ho R T K and Yu C P 2000
Stereotactic dose planning system used in Leksell GammaKnife
model-B: EGS4 Monte Carlo versus GafChromic films
MD-55
{\it Appl. Radiat. Isot.} {\bf 53} 427-30

\item[] Cheung J Y C, Yu K N, Yu C P and Ho R T K 1998
Monte Carlo calculation of single-beam dose profiles used in a gamma
knife treatment planning system
{\it Med. Phys.} {\bf 25} 1673-5

\item[] Cheung J Y C, Yu K N, Yu C P and Ho R T K 2001
Dose distributions at extreme irradiation depths of gamma knife
radiosurgery: EGS4 Monte Carlo
calculations 
{\it Appl. Radiat. Isot.} {\bf 54} 461-5

\item[] Elekta 1992 {\it Leksell Gamma Unit-User's Manual} (Stockholm:
Elekta Instruments AB)

\item[] Elekta 1996 {\it Leksell GammaPlan Instructions for Use for
Version 4.0-Target Series} (Geneva: Elekta)

\item[] Hartmann G H, Lutz W, Arndt J, Ermakov I, Podgorsak E B, Schad
L, Serago C and Vatinisky S M (ed) 1995 Quality assurance program on
stereotactic radiosurgery {\it Report from a Quality Assurance Task
Group} (Berlin: Springer)

\item[] Moskvin V, DesRosiers C, Papiez L, Timmerman R, Randall M and
DesRosiers P 2002 
Monte Carlo simulation of the Leksell GammaKnife:
I. Source modelling and calculations in homogeneous media 
{\it Phys. Med. Biol.} {\bf 47} 1995-2011
 
\item[] Salvat F, Fern\'andez-Varea J M, Acosta E and Sempau J 2001
{\it PENELOPE, a code system for Monte Carlo simulation of
electron and photon transport} (Paris: NEA-OECD)

\item[] Semapu J, S\'anchez-Reyes A, Salvat F, Oulad ben Tahar H,
Jiang S B and Fern\'andez-Varea J M 2001 Monte Carlo simulation of
electron beams from an accelerator head using PENELOPE {\it
Phys. Med. Biol.} {\bf 46} 1163-86

\item[] Solberg T D, DeMarco J J, Holly F E, Smathers J B and DeSalles
A A F 1998 Monte Carlo treatment planning for stereotactic
radiosurgery {\it Radiother. Oncol.} {\bf 49} 73-84

\item[] Wu A 1992 Physics and dosimetry of the gamma knife
{\it Neurosurg. Clin. N. Am.} {\bf 3} 35-50

\item[] Wu A, Lindner G, Maitz A, Kalend A, Lunsfond L D, Flickinger J C 
and Bloomer W D 1990 
Physics of gamma knife approach on convergent beams in stereotactic
radiosurgery  
{\it Int. J. Radiation Oncology Biol. Phys.} {\bf 18},
941-9

\item[] Xiaowei L and Chunxiang Z 1999 
Simulation of dose distribution irradiation by the Leksell Gamma Unit 
{\it Phys. Med. Biol.} {\bf 44} 441-5

\endrefs

\newpage

\end{document}